\documentclass{ws-ijmpa}

\begin{document}

\markboth{JULIO OLIVA, DAVID TEMPO AND RICARDO TRONCOSO}
{Static Wormholes In Vacuum For Conformal Gravity}

%
\catchline{}{}{}{}{}
%

\title{STATIC WORMHOLES IN VACUUM\\FOR CONFORMAL GRAVITY}

\author{JULIO OLIVA}

\address{Centro de Estudios Cient\'{\i}ficos (CECS), Casilla 1469, Valdivia, Chile.\\
Instituto de F\'{\i}sica, Universidad Austral de Chile, Casilla 567, Valdivia, Chile.}

\author{DAVID TEMPO}

\address{Centro de Estudios Cient\'{\i}ficos (CECS), Casilla 1469, Valdivia, Chile. \\
Departamento de F\'{\i}sica, Universidad de Concepci\'{o}n, Casilla
160-C, Concepci\'{o}n, Chile.\\
Physique th\'{e}orique et math\'{e}matique, Universit\'{e} Libre de Bruxelles,
ULB Campus Plaine C.P.231, B-1050 Bruxelles, Belgium.}

\author{RICARDO TRONCOSO}

\address{Centro de Estudios Cient\'{\i}ficos (CECS), Casilla 1469, Valdivia, Chile.\\
Centro de Ingenier\'{\i}a de la Innovaci\'{o}n del CECS (CIN), Valdivia, Chile.}

\maketitle

\begin{history}
\end{history}

\begin{abstract}
A static spherically symmetric wormhole solution for conformal
gravity in vacuum is found. The solution possesses a single integration
constant which determines the size of the neck connecting two static
homogeneous universes of constant spatial curvature. Time runs at different
rates on each side of the neck, and depending on the value of the parameter,
the wormhole can develop a cosmological horizon only at one side. It is shown that
 the wormholes correspond to the matching
of different Einstein spacetimes by means of improper conformal
transformations.

\end{abstract}

\ccode{PACS numbers: 04.20.Jb, 04.50.Kd}

If one wonders about the possibility of realizing static wormholes in vacuum
in four dimensions, one knows that it should be performed within some scenario
beyond General Relativity (see e.g. Ref. \refcite{VisserBOOK}). Experience gathered
from extensions of General Relativity in diverse dimensions\cite{DOTWorm}$^{-}$\cite{CC} suggests that one of the key
features allowing this possibility is the relaxation of the asymptotic
conditions for the metric. An interesting theory possessing this last feature
is conformal gravity (see e.g. Refs. \refcite{FT},\refcite{BHn},\refcite{BW}), whose
Lagrangian can be written as the square of the Weyl Tensor, so that the action
reads%
\begin{equation}
I=\alpha\int d^{4}x\sqrt{-g}\ C^{\alpha\beta\mu\nu}C_{\alpha\beta\mu\nu}\ .
\end{equation}
Apart from being invariant under diffeomorphisms, this action is also
invariant under local rescalings of the metric given by $g_{\mu\nu}%
\rightarrow\Omega^{2}(x)g_{\mu\nu}$, and the field equations are given by the
vanishing of the Bach tensor, i.e.,
\begin{equation}
\left(  \nabla^{\mu}\nabla^{\nu}+\frac{1}{2}R^{\mu\nu}\right)  C_{\hspace
{0.06in}\mu\beta\nu}^{\alpha}=0\ .
\end{equation}
It is simple to verify that any Einstein space with cosmological constant is a
solution of the theory in vacuum.

Conformal gravity was intensively studied in the past and it has been shown to
be renormalizable\cite{Stelle} -- Note that the coupling $\alpha$ is
dimensionless. Nevertheless, the theory possesses fourth-order field equations
for the metric, so that it has ghosts as it is generically expected. In the
context we are interested in, regardless the theory is suitable or not as the
ultimate one to describe gravity, it certainly deserves to be studied.

It is also worth pointing out that it has been recently shown that for smooth
matter distributions, wormholes that do not violate the weak energy condition
near the throat can exist in conformal gravity.\cite{LOBOWEYL} For our purposes, it is useful mentioning that the most general spherically
symmetric solution of conformal gravity (see e.g. Ref. \refcite{Riegert}) possesses
a relaxed asymptotic behavior as compared with General Relativity. Thus, for
the reasons explained above, the door is open to look for wormholes in vacuum
within this theory.

It can be seen that conformal gravity admits the following static spherically
symmetric wormhole solution in vacuum, whose metric reads%

\begin{equation}
ds^{2}=-\left(  1+a^{2}\tanh\rho  \right)  dt^{2}+\frac
{d\rho^{2}}{1+a^{2}\tanh\rho  }+l_{0}^{2}\cosh^{2}\rho
\ d\Omega^{2}\ ,\label{spherical-wormhole-vacuum}%
\end{equation}
where $d\Omega^{2}$ stands for the line element of $S^{2}$, and the range of
the radial coordinate is given by $-\infty<\rho<\infty$. The wormhole
possesses a single integration constant $a$ that parameterizes the radius of
the neck given by $l_{0}^{2}=\left(  3a^{4}+1\right)  ^{-1/2}$, being located
at $\rho=0$. The wormhole connects two static homogeneous universes of
constant spatial curvature with different radii, as it can be seen from the
asymptotic behavior of the curvature%

\begin{equation}
R_{\ \ kl}^{mn}\underset{\rho\rightarrow\pm\infty}{\longrightarrow}-\left(
1\pm a^{2}\right)  \delta_{\ \ kl}^{mn}\ ,
\end{equation}
and time runs at different rates at each side of the neck, since%
\begin{equation}
g_{tt}\underset{\rho\rightarrow\pm\infty}{\longrightarrow}-\left(  1\pm
a^{2}\right)  \ .
\end{equation}

The case of $a^{2}\rightarrow-a^{2}$ is obtained from reflexion symmetry of
the radial coordinate.

For $a^{2}<1$ the wormhole interpolates between static universes with spatial
geometries given by hyperbolic spaces of radii $\left(  1\pm a^{2}\right)
^{-1/2}$. In the case of $a=0$ the metric acquires a simple form, given by%
\begin{equation}
ds^{2}=-dt^{2}+d\rho^{2}+\cosh^{2}\rho\ d\Omega^{2}\ ,
\end{equation}
which at both sides is asymptotically locally described by $R\times H_{3}$,
and it can be regarded as \textquotedblleft the groundstate".
For $a=1$ the wormhole
interpolates between flat space and a static universe with spatial geometry
given by a hyperbolic space of radius $2^{-1/2}$. In the case of $a^{2}>1$ the
wormhole develops a cosmological horizon at one side of the neck, located at
$\rho=\rho_{c}=-\tanh^{-1}\left(  1/a^{2}\right)  $, and interpolates between
an Einstein Universe ($R\times S^{3}$) for $\rho\rightarrow-\infty$, and
$R\times H_{3}$.

The causal structure of the wormhole coincides with the one of Minkowski
spacetime in two dimensions for $a^{2}\leq1$. For $a^{2}>1$ the wormhole shares the same causal
structure with two-dimensional Rindler spacetime, as shown in Fig. 1a. \\
It can also be shown that the wormholes in vacuum generically correspond to
the matching of different Einstein spacetimes at infinity by means of improper
conformal transformations. A particularly interesting case corresponds to $a^2>1$ since the
wormhole is conformally related to the matching of the patch covering the region within the event horizon and infinity of the Schwarzschild-de Sitter (SdS) metric
with positive mass with two additional patches obtained from the region beyond the cosmological horizon of
the Schwarzschild-de Sitter metric with negative mass. This is depicted in Fig. 1b.
Note that in General Relativity the matching of different spaces through the boundary of their corresponding conformal compactifications
has no sense, since the proper distance to pass from one patch to the other diverges. Nonetheless, it is worth pointing out that
within the context of conformal gravity this kind of matching can be performed by means of an improper conformal transformation, i.e., a conformal
transformation that vanishes at the matching surface. In this way, the proper distance required to pass through a pair of points located at each side of the matching surface, given by $ds^2=\Omega^2 ds_E^2$, becomes finite.

It can also be seen that the wormhole (\ref{spherical-wormhole-vacuum}) is conformally related by patches to the spherically symmetric
 solution of conformal gravity found in Refs. \refcite{Bach} and \refcite{Riegert}.
\begin{figure}[pb]
\centerline{\psfig{angle=-90,file=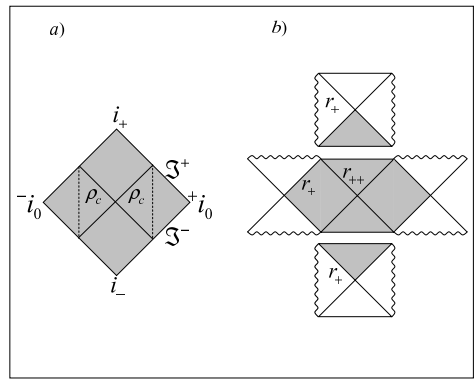,width=6.5cm}}
\vspace*{8pt}
\caption{(a) Corresponds to the causal structure of the wormhole with $a>1$. (b) Shows how this wormhole
 is constructed from the conformal matching of two SdS spacetimes with positive and negative masses.}
\end{figure}
This feature also holds for the
generalization of the wormholes presented here, having necks of genus greater than one. The extension of the wormhole (\ref{spherical-wormhole-vacuum}) with radial
electric or magnetic fields can also be found, and it turns out to have
electric or magnetic \textquotedblleft charge without charge". For further details we refer to Ref. \refcite{OTT1}.
As an ending remark, it is worth mentioning that the definition of mass in
conformal gravity is a very subtle issue, which is not free of controversy.\cite{BHS}$^{-}$\cite{D1} Thus, in order to have a suitable analysis of the
solutions, the construction of finite conserved charges written as suitable
surface integrals for conformal gravity should be addressed.

\section*{Acknowledgments}

We thank the organizers of the ``7th Alexander Friedmann International Seminar on Gravitation and Cosmology'' for the nice atmosphere provided at the meeting. Special thanks to Gast\'{o}n Giribet, Carlos
Kozameh, Claude LeBrun and Roberto Troncoso for very useful comments. This work was
partially funded by FONDECYT grants 1095098, 1061291, 1071125, 1085322, 3085043; David Tempo thanks CONICYT and Escuela de Graduados of the Universidad de Concepci\'{o}n for financial support. The Centro
de Estudios Cient\'{\i}ficos (CECS) is funded by the Chilean Government
through the Millennium Science Initiative and the Centers of Excellence Base
Financing Program of CONICYT. CECS is also supported by a group of private
companies which at present includes Antofagasta Minerals, Arauco, Empresas
CMPC, Indura, Naviera Ultragas and Telef\'{o}nica del Sur. CIN is funded by
CONICYT and the Gobierno Regional de Los R\'{\i}os.


\end{document}